\documentstyle[preprint,aps,epsf,floats]{revtex} 
\begin{document}

{\tighten
%%\twocolumn[\hsize\textwidth\columnwidth\hsize\csname@twocolumnfalse\endcsname

\preprint{\vbox{\hbox{CALT-68-2043}
                \hbox{hep-ph/9603314} 
		\hbox{\footnotesize DOE RESEARCH AND}
		\hbox{\footnotesize DEVELOPMENT REPORT} }}
 
\title{Implications of the $B\to X\,\ell\,\bar\nu_\ell$ lepton spectrum \\
  for heavy quark physics }
 
\author{Martin Gremm, Anton Kapustin, Zoltan Ligeti and Mark B. Wise}

\address{California Institute of Technology, Pasadena, CA 91125}

\maketitle 
\widetext

\begin{abstract}%
The shape of the lepton spectrum in inclusive semileptonic $B\to
X\,\ell\,\bar\nu_\ell$ decay is sensitive to matrix elements of the heavy quark
effective theory, $\bar\Lambda$ and $\lambda_1$.  From CLEO data we extract
$\bar\Lambda=0.39\pm0.11\,$GeV and $\lambda_1=-0.19\pm0.10\,{\rm GeV}^2$, where
the uncertainty is the $1\sigma$ statistical error only.  Systematic
uncertainties are discussed.  These values for $\bar\Lambda$ and $\lambda_1$
are used to determine $|V_{cb}|$ and the $\overline{\rm MS}$ bottom and charm 
quark masses.  We discuss the theoretical uncertainties related to order
$(\Lambda_{\rm QCD}/m_b)^3$ effects and higher orders in the perturbative
expansion.  
\end{abstract}

}%end tighten
\newpage 
%%\pacs{12.15.Ff, 12.15.Hh, 12.39.Hg, 13.20.He}
%%]\narrowtext

The operator product expansion (OPE) shows that in the limit
$m_b\gg\Lambda_{\rm QCD}$ inclusive semileptonic $B$ decay rates are equal to
the perturbative $b$ quark decay rates \cite{CGG}.  Experimental study of such
decays provide measurements of fundamental parameters of the standard model,
such as the CKM angles $|V_{cb}|$, $|V_{ub}|$, and the bottom and charm quark
masses.

To obtain precise theoretical predictions for inclusive semileptonic $B$
decays, it is important to be able to compute nonperturbative effects
suppressed by powers of $\Lambda_{\rm QCD}/m_b$.  There are no nonperturbative
corrections at order $\Lambda_{\rm QCD}/m_b$, and the corrections of order
$(\Lambda_{\rm QCD}/m_b)^2$ are characterized by only two matrix elements
\cite{incl,MaWi} 
\begin{equation}
\lambda_1 = {1\over2m_M}\, \langle M(v) \,|\, \bar h_v\, (iD)^2\, 
  h_v \,|\, M(v)\rangle \,,
\end{equation}
and 
\begin{equation}
\lambda_2 = {1\over2d_Mm_M}\, \langle M(v) \,|\, \bar h_v\, {g\over2}\,
  \sigma_{\mu\nu}\, G^{\mu\nu}\, h_v \,|\, M(v)\rangle \,,
\end{equation}
where $M$ is a pseudoscalar or vector meson containing a heavy quark $Q$, 
and $h_v$ is the heavy quark field in the effective theory \cite{eft} with 
velocity $v$.  $d_M=3,-1$ for pseudoscalar or vector mesons, respectively.
The decay rates also depend on the quark masses, which can be expressed in 
terms of the heavy meson masses and the parameters $\lambda_1$, $\lambda_2$ 
and $\bar\Lambda$, where
\begin{equation}\label{masses}
m_M = m_Q + \bar\Lambda - {\lambda_1+d_M\lambda_2\over2m_Q} + \ldots \,.
\end{equation}

The matrix element $\lambda_2$ is then determined from the measured $B^*-B$
mass splitting, $\lambda_2\simeq0.12\,{\rm GeV}^2$.  The quantity $\bar\Lambda$
\cite{Luke} also sets the scale for the deviation of the exclusive $B\to
D^{(*)}\,\ell\,\bar\nu_\ell$ decay form-factors from the Isgur-Wise function
\cite{HQS}.  The analogue of $\bar\Lambda$ in the baryon sector,
$\bar\Lambda_\Lambda=m_{\Lambda_b}-m_b+\ldots$, describes all $\Lambda_{\rm
QCD}/m_Q$ corrections to $\Lambda_b\to\Lambda_c\,\ell\,\bar\nu_\ell$ decays
\cite{GGW}, and is related to $\bar\Lambda$ via
$\bar\Lambda_\Lambda=\bar\Lambda+m_{\Lambda_b}-m_B+\ldots$.

To carry out accurate calculations it is crucial to have reliable
determinations of $\bar\Lambda$ and $\lambda_1$.  In the past, these quantities
have been estimated using models of QCD \cite{models}, and extracting them from
experimental data was attempted \cite{LuSa,YoZ,AnZ,FLSn}.  Sum rules were also
derived to constrain $\lambda_1$ \cite{BSUV}, however, perturbative corrections
weaken these constraints \cite{BAZM}.  In this letter we extract $\bar\Lambda$
and $\lambda_1$ from the shape of the inclusive $B\to X\,\ell\,\bar\nu_\ell$
lepton spectrum, and also translate our results into a determination of
$|V_{cb}|$, and the $\overline{\rm MS}$ masses $\overline{m}_b(m_b)$, and
$\overline{m}_c(m_c)$.

The CLEO Collaboration has measured the inclusive $B\to X\,\ell\,\bar\nu_\ell$
lepton spectrum both by demanding only one charged lepton tag \cite{1tag}, and
using a double tagged data sample \cite{2tag} where the charge of a high
momentum lepton determines whether the other lepton in the event comes directly
from semileptonic $B$ decay (primary) or from the semileptonic decay of a $B$
decay product charmed hadron (secondary).  The single tagged data sample has
smaller statistical errors, but it is significantly contaminated by secondary
leptons below about $1.5\,$GeV.  For our analysis we use the data as tabulated
in Ref.~\cite{RoyPhD}.

The OPE for the lepton spectrum in semileptonic $B$ decay does not reproduce
the physical lepton spectrum point-by-point near maximal lepton energy.  Near
the endpoint, comparison with experimental data can only be made after
sufficient smearing, or after integrating over a large enough region.  The
minimal size of this region was estimated to be around $300-500\,$MeV
\cite{MaWi}.  This, and the fact that the experimental measurement of the
lepton spectrum is precise and model independent only above about $1.5\,$GeV,
impose a limitation on what quantities can be reliably predicted and compared
with data.  On the one hand, we want to find observables sensitive to
$\bar\Lambda$ and $\lambda_1$; on the other hand, we want the deviations from
the $b$ quark decay prediction to be small, so that the contributions from even
higher dimension operators in the OPE are not too important.  The observables
we use should not depend on $|V_{cb}|$.  Thus we consider
\begin{equation}\label{R12def}
R_1 = {\displaystyle \int_{1.5\,{\rm GeV}} E_\ell\,
  {{\rm d}\Gamma\over{\rm d}E_\ell}\, {\rm d}E_\ell \over \displaystyle
  \int_{1.5\,{\rm GeV}} {{\rm d}\Gamma\over{\rm d}E_\ell}\, 
  {\rm d}E_\ell }\,, \qquad
R_2 = {\displaystyle \int_{1.7\,{\rm GeV}} 
  {{\rm d}\Gamma\over{\rm d}E_\ell}\, {\rm d}E_\ell \over \displaystyle
  \int_{1.5\,{\rm GeV}} {{\rm d}\Gamma\over{\rm d}E_\ell}\, 
  {\rm d}E_\ell }\,.
\end{equation}

Before comparing the experimental data with the theoretical predictions for
$R_{1,2}$, derived from the OPE and QCD perturbation theory, the following
corrections have to be included: ($i$) electromagnetic radiative correction;
($ii$) effects of boost into the lab frame; ($iii$) smearing due to detector
momentum resolution.  To take ($i$) into account, following the CLEO analysis,
we used the resummed photon radiation corrections as given in Ref.~\cite{AtMa}. 
These corrections to $R_{1,2}$ have very little sensitivity to subleading
logarithms.  To determine the corrections due to ($ii$), we assume that the $B$
mesons are monoenergetic, with energy $m_{\Upsilon(4S)}/2$ (the effect of the
$4\,$MeV spread in the center of mass energy is negligible).  We found that the
smearing due to the CLEO-II detector momentum resolution \cite{CLEOII}, and the
$50\,$MeV binning of the data has a negligible effect on $R_{1,2}$.

Including the leading nonperturbative corrections of order $(\Lambda_{\rm
QCD}/m_b)^2$ \cite{incl,MaWi} and the order $\alpha_s$ corrections
\cite{alphas}, the theoretical expressions for $R_{1,2}$ are
\begin{mathletters}\label{R12exp}
\begin{eqnarray}
R_1 &=& 1.8059 - 0.309\,{\bar\Lambda\over\overline{m}_B} 
  - 0.35\,{\bar\Lambda^2\over\overline{m}_B^2}  
  - 2.32\,{\lambda_1\over\overline{m}_B^2} 
  - 3.96\,{\lambda_2\over\overline{m}_B^2}  
  - {\alpha_s\over\pi} \left(0.035+0.07\,{\bar\Lambda\over\overline{m}_B}\right)
  \nonumber\\
&+& \left|{V_{ub}\over V_{cb}}\right|^2 
  \left( 1.33 - 10.3\,{\bar\Lambda\over\overline{m}_B} \right)  
  - \left(0.0041-0.004\,{\bar\Lambda\over\overline{m}_B}\right)  
  + \left(0.0062+0.002\,{\bar\Lambda\over\overline{m}_B}\right) ,\label{R1exp}\\
R_2 &=& 0.6581 - 0.315\,{\bar\Lambda\over\overline{m}_B} 
  - 0.68\,{\bar\Lambda^2\over\overline{m}_B^2}  
  - 1.65\,{\lambda_1\over\overline{m}_B^2} 
  - 4.94\,{\lambda_2\over\overline{m}_B^2}  
  - {\alpha_s\over\pi} \left(0.039+0.18\,{\bar\Lambda\over\overline{m}_B}\right)
  \nonumber\\
&+& \left|{V_{ub}\over V_{cb}}\right|^2 
  \left( 0.87 - 3.8\,{\bar\Lambda\over\overline{m}_B} \right) 
  - \left(0.0073+0.005\,{\bar\Lambda\over\overline{m}_B}\right)  
  + \left(0.0021+0.003\,{\bar\Lambda\over\overline{m}_B}\right) , \label{R2exp}
\end{eqnarray}
\end{mathletters}%
with $R_1$ in GeV.  We have defined the spin-averaged $B$ meson mass,
$\overline{m}_B=(m_B+3m_{B^*})/4$.  The terms in the last two parentheses
in each of eqs.~(\ref{R1exp}) and (\ref{R2exp}) represent the effect of the
electromagnetic radiative correction, and the effect of the boost into the lab
frame, respectively.  The parts of these corrections proportional to
$\lambda_{1,2}$ are negligible.  Eqs.~(\ref{R12exp}) correspond to electrons;
for muons the electromagnetic correction is smaller.  Corrections to
eqs.~(\ref{R12exp}) of higher order in $\alpha_s$ and $\Lambda_{\rm QCD}/m_b$ 
will be discussed later.  Even though there are no nonperturbative corrections
to $R_{1,2}$ of order $\Lambda_{\rm QCD}/m_b$, eqs.~(\ref{R12exp}) contain
terms proportional to $\bar\Lambda/\overline{m}_B$.  These arise since we
reexpressed the heavy quark masses in terms of hadron masses, using
$m_Q=\overline{m}_M-\bar\Lambda+\lambda_1/(2\overline{m}_M)$.

To compare the above theoretical expressions with data, we need to discuss 
the experimental uncertainties.  We use the single tagged data to extract
$R_{1,2}$, and correct for the effects of secondary leptons \cite{VC} using 
the double tagged data.  The central values are $R_1=1.7831\,$GeV and 
$R_2=0.6159$ (the corrections from the secondaries are 0.0001 and 0.0051 
respectively), while the correlation matrix of the statistical errors 
\cite{RoyPhD} is
\begin{equation}
V(R_1,R_2) = \pmatrix{ 3.8\times10^{-6} & 6.0\times10^{-6} \cr
  6.0\times10^{-6} & 1.7\times10^{-5} } .
\end{equation}
The largest part of these uncertainties is due to the errors in the
secondary lepton spectrum from the double tagged data.  Estimating the
systematic errors is more complicated.  These uncertainties in the lepton
spectrum can be divided into two classes: there are additive corrections, like
backgrounds that are subtracted from the data; and there are multiplicative
corrections, like those in efficiencies.  The total systematic uncertainty in
the CLEO measurement of the semileptonic $B$ decay branching fraction is about
2\%.  However, only a small fraction of these uncertainties affect the shape of
the lepton spectrum above $1.5\,$GeV \cite{Roy}.  In this region the
uncertainties in the backgrounds are small, and the efficiencies have fairly
flat momentum dependences.  While the uncertainties in the electron
identification and in the tracking efficiencies are the dominant sources of
systematic error in the semileptonic $B$ branching fraction, they are expected
to affect $R_{1,2}$ at a much smaller level.  We estimate that the systematic
uncertainties in $R_{1,2}$ are not larger than the statistical errors
\cite{Roy}, although a complete analysis of these can only be carried out by
the CLEO Collaboration.  For this reason, and since the statistical errors can
be included into our analysis exactly, the experimental uncertainties we shall
quote will be the statistical ones only.

The comparison of the theoretical predictions in eqs.~(\ref{R12exp}) with the
CLEO data is shown in Fig.~1.  The steeper band is the constraint from $R_2$,
while the hatched one is that from $R_1$.  The widths of the bands represent
the $1\sigma$ statistical errors, while the ellipse shows the $1\sigma$ allowed
region in $\{\bar\Lambda,\lambda_1\}$, after correlations between $R_1$ and
$R_2$ are taken into account.  This region corresponds to
$\bar\Lambda=0.39\pm0.11\,$GeV and $\lambda_1=-0.19\pm0.10\,{\rm GeV}^2$.  
The $1\sigma$ allowed region in Fig.~1 lies outside but not far from the 
region allowed by a recent analysis based on moments of the hadron spectrum 
\cite{FLSn}.  

\begin{figure}[tb]  %% 8.5truecm for twocolumn; 11truecm for preprint
\centerline{\epsfxsize=8.5truecm \epsfbox{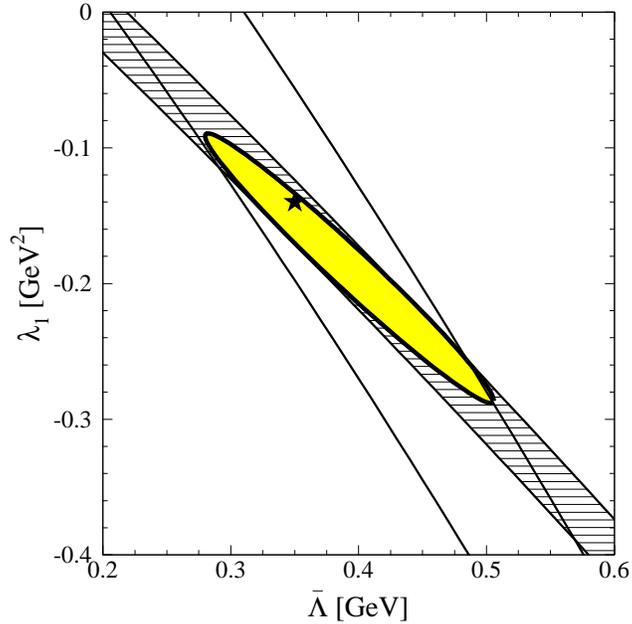}}
\caption[1]{Allowed regions in the $\bar\Lambda-\lambda_1$ plane
for $R_1$ and $R_2$.  The bands represent the $1\sigma$ statistical 
errors, while the ellipse is the allowed region taking correlations
into account.  The star denotes where the order $(\Lambda_{\rm QCD}/m_b)^3$ 
corrections discussed in the text would shift the center of the ellipse. }
\end{figure}

In Fig.~1 we set $|V_{ub}/V_{cb}|=0.08$.  The extraction of this value is model
dependent, and therefore has considerable uncertainty.  If $|V_{ub}/V_{cb}|=0.1$
then the center of the ellipse in Fig.~1 would move to $\bar\Lambda=0.42\,$GeV
and $\lambda_1=-0.19\,{\rm GeV}^2$.  We used $\alpha_s=0.22$, corresponding to
the subtraction scale $m_b$.  The sensitivity of our results to this choice of
scale is weak; changing $\alpha_s$ to 0.35 moves the central values to
$\bar\Lambda=0.36\,$GeV and $\lambda_1=-0.18\,{\rm GeV}^2$.

To plot Fig.~1 we used the data corresponding to electrons only, as we suspect
that the systematic uncertainties in the (single tagged) muon data may be
larger (for example, the muon detection efficiency is strongly energy dependent
below $2\,$GeV).  The latter data set, nevertheless, yields a consistent
determination of $\bar\Lambda$ and $\lambda_1$, giving central values
$\bar\Lambda=0.43\,$GeV and $\lambda_1=-0.21\,{\rm GeV}^2$ (to subtract 
secondaries we used the double tagged electron data).

Theoretical uncertainties in our determination of $\bar\Lambda$ and $\lambda_1$
originate from the reliability of quark-hadron duality at the scales
corresponding to the limits in the integrals defining $R_{1,2}$, from order
$(\Lambda_{\rm QCD}/m_b)^3$ corrections, and from higher order perturbative
corrections.  Concerning duality, note that $E_\ell\geq1.5\,$GeV and $1.7\,$GeV
(in the lab frame) correspond to summing over hadronic states $X$ with masses
below $3.6\,$GeV and $3.3\,$GeV, respectively.  These scales are likely to be
large enough to trust the OPE locally.  This is supported by the fact that a
modified ratio that differs from $R_2$ only in that the integration limit in the
numerator is changed from $1.7\,$GeV to $1.8\,$GeV yields a parallel band that
overlaps with that corresponding to $R_2$.  Using this variable and
$R_1$, the central values for $\bar\Lambda$ and $\lambda_1$ become
$\bar\Lambda=0.47\,$GeV and $\lambda_1=-0.26\,{\rm GeV}^2$.  (The assumption of
local duality becomes less reliable using $1.8\,$GeV.)  For higher moments
\cite{Volo} theoretical uncertainties increase, and they are sensitive to an
almost identical combination of $\bar\Lambda$ and $\lambda_1$ as the first
moment, $R_1$.  For example, the normalized second moment (with
$E_\ell>1.5\,$GeV) gives a band that overlaps with that from $R_1$, and together
with $R_2$ yields the central values $\bar\Lambda=0.39\,$GeV and
$\lambda_1=-0.19\,{\rm GeV}^2$.  Perturbative corrections of order $\alpha_s^2$
and order $\alpha_s(\Lambda_{\rm QCD}/m_b)^2$ have not been computed.  The order
$\alpha_s^2$ corrections are likely to decrease the magnitudes of $\bar\Lambda$
and $\lambda_1$.

The recently calculated order $(\Lambda_{\rm QCD}/m_b)^3$ correction to the
differential decay rate \cite{GrKa} is parametrized by two matrix elements,
$\rho_1$ and $\rho_2$,  and time ordered products of local operators.  These
also modify the relation between the quark and hadron masses.  Neglecting
$\rho_2$ (that is expected to be small \cite{BSUV}) and the time ordered
products, the ellipses in eq.~(\ref{masses}) is replaced by $\rho_1/(2m_Q)^2$;
and $\lambda_2$ is determined by
$m_{B^*}-m_B=(2\lambda_2/\overline{m}_B)(1+\bar\Lambda/\overline{m}_B)$.  

To estimate the possible size of these $(\Lambda_{\rm QCD}/m_b)^3$ corrections,
we calculated the effect of the terms proportional to $\bar\Lambda^3$,
$\bar\Lambda\lambda_{1,2}$ and $\rho_1$ on the central values of $\bar\Lambda$
and $\lambda_1$.  The numerical value of $\rho_1$ can be estimated using the
vacuum saturation approximation,
$\rho_1=(2\pi\alpha_s/9)\,m_B\,f_B^2\simeq(300\,{\rm MeV})^3$ (we agree with
\cite{BSUV} and disagree with \cite{Mannel} on this result).  Reexpanding
$R_{1,2}$ to order $(\Lambda_{\rm QCD}/m_b)^3$, we get
\begin{mathletters}\label{R12shift}
\begin{eqnarray}
\delta R_1 &=& -(0.4\bar\Lambda^3 + 5.7\bar\Lambda\lambda_1 
  + 6.8\bar\Lambda\lambda_2 + 7.7\rho_1)/ \overline{m}_B^3 \,, \\
\delta R_2 &=& -(1.5\bar\Lambda^3 + 7.1\bar\Lambda\lambda_1 
  + 17.5\bar\Lambda\lambda_2 + 1.8\rho_1)/ \overline{m}_B^3 \,.
\end{eqnarray}
\end{mathletters}%
Solving the constraint imposed by the experimental data, our central values
(the center of the ellipse in Fig.~1) are moved to the point denoted with a
star in Fig.~1, $\bar\Lambda=0.35\,$GeV and $\lambda_1=-0.15\,{\rm GeV}^2$.  
We think that the terms displayed in eqs.~(\ref{R12shift}) are likely to be the
dominant order $(\Lambda_{\rm QCD}/m_b)^3$ corrections.  Uncertainties in the
values of $\rho_2$ and in the time ordered products are expected to have a
numerically smaller effect on $R_{1,2}$.  Therefore, the $(\Lambda_{\rm
QCD}/m_b)^3$ corrections will probably reduce the magnitudes of both
$\bar\Lambda$ and $\lambda_1$.

$\bar\Lambda$ is not a physical quantity, and has a ``renormalon ambiguity" of
order $\Lambda_{\rm QCD}$ \cite{renorm}.  The perturbative expression for the
semileptonic decay rate in terms of the $b$ quark pole mass $m_b$ is not Borel
summable, and neither is the perturbative expansion of the $\overline{\rm MS}$
mass $\overline{m}_b(m_b)$ in terms of $m_b$.  These ambiguities cancel if
$\bar\Lambda$ (or equivalently the $b$ quark pole mass) extracted from the
differential semileptonic decay rate is used to get the $\overline{\rm MS}$
mass.  Consequently one can arrive at a meaningful prediction for
$\overline{m}_b(m_b)$.  It is fine to introduce unphysical quantities like
$\bar\Lambda$, as long as one works consistently to a given order of QCD
perturbation theory and the expansion in inverse powers of the heavy quark
masses.  Since the final results one considers always involve relations between
physically measurable quantities, any ``renormalon ambiguities" arising from
the bad behavior of the QCD perturbation series at large orders cancel out
\cite{rencan}.  

In summary, using CLEO data on the lepton spectrum from inclusive
semileptonic $B\to X\,\ell\,\bar\nu_\ell$ decay, we obtained the values of
the heavy quark effective theory matrix elements,
$\bar\Lambda=0.39\pm0.11\,$GeV and $\lambda_1=-0.19\pm0.10\,{\rm GeV}^2$, 
where the uncertainty corresponds to the $1\sigma$ statistical error.  
These imply at order $\alpha_s$ that 
\begin{equation}\label{Vcb}
|V_{cb}|= 0.041 \left( {{\rm Br}(B\to X_c\,\ell\,\bar\nu_\ell)\over0.105}\,
  {1.54\,{\rm ps}\over\tau_B}\right)^{1/2} .
\end{equation}
The largest uncertainty ($\sim4$\%) in this determination of $|V_{cb}|$ is due
to higher order corrections in $\alpha_s$ ({\it e.g.}, the order
$\alpha_s^2\beta_0$ correction to the inclusive semileptonic $B$ decay rate
\cite{LSW} increases the factor 0.041 in eq.~(\ref{Vcb}) to 0.042).  

The difference between the bottom and charm quark pole masses is free of
renormalon ambiguities (at order $\Lambda_{\rm QCD}$).  We find
$m_b-m_c=3.37\pm0.02\,$GeV, where the uncertainty is the $1\sigma$ statistical
error.  The $\overline{\rm MS}$ quark masses are related to the pole mass
via $\overline{m}_Q(m_Q)=m_Q[1-4\alpha_s(m_Q)/(3\pi)+\ldots]$.  Using this and
our values for $\bar\Lambda$ and $\lambda_1$, we obtain at order $\alpha_s$
\begin{equation}\label{msmasses}
\overline{m}_b(m_b) = 4.45\,{\rm GeV}  \,, \qquad
\overline{m}_c(m_c) = 1.28\,{\rm GeV} \,.
\end{equation}
Order $\alpha_s^2$ terms in the relation between $m_Q$ and 
$\overline{m}_Q(m_Q)$ reduce these $\overline{\rm MS}$ heavy quark masses
by about $230\,$MeV and $180\,$MeV, respectively.  Of course, it is not
consistent to use these corrections as order $\alpha_s^2$ terms have not
been included into our determination of $\bar\Lambda$ and $\lambda_1$
through $R_{1,2}$.  Recent determinations of the $\overline{\rm MS}$
$b$ quark mass using lattice QCD are $\overline{m}_b(m_b)=4.17\pm0.06\,$GeV
and $\overline{m}_b(m_b)=4.0\pm0.1\,$GeV \cite{lattice}.  Given the
uncertainties in eq.~(\ref{msmasses}) from order $\alpha_s^2$ effects
(and since the lattice results only contain the one-loop matching between 
lattice and continuum perturbation theory), our result for 
$\overline{m}_b(m_b)$ is consistent with these.

The bands in Fig.~1 corresponding to $R_{1,2}$ are almost parallel.  Hence,
even small corrections to eqs.~(\ref{R12exp}) can significantly affect our
determination of $\bar\Lambda$ and $\lambda_1$.  It would be useful to have
constraints that depend on very different combinations of $\bar\Lambda$ and
$\lambda_1$.  The moments of the photon spectrum in inclusive $B\to
X_s\,\gamma$ decay \cite{AnZ} can provide such information.

Some improvements on the analysis in this paper are possible.  The dominant
(order $\alpha_s^2\beta_0$) part of the order $\alpha_s^2$ corrections should be
readily calculable \cite{LSW}.  The order $\alpha_s(\Lambda_{\rm QCD}/m_b)^2$
corrections would set the scale for $\lambda_2$, and could also yield sizable
corrections to the coefficients of $\lambda_1$ in eqs.~(\ref{R12exp}).  A better
knowledge of the order $(\Lambda_{\rm QCD}/m_b)^3$ corrections and a model
independent determination of $|V_{ub}|$ would also reduce the uncertainties. 
The systematic experimental errors on $R_{1,2}$ should be evaluated by the CLEO
Collaboration, and hopefully in the future both the systematic and the
statistical errors can be reduced by a larger sample of double tagged
semileptonic decays.  Such a sample may also permit the use of variables
analogous to $R_{1,2}$, but with smaller lower limits of integration, thereby
reducing uncertainties associated with the validity of quark-hadron duality.

We thank R. Wang, J. Urheim and A. Weinstein for their assistance.
This work was supported in part by the U.S.\ Dept.\ of Energy under Grant no.\
DE-FG03-92-ER~40701.  A.K.\ was also supported by the Schlumberger Foundation.

%%\newpage
{\tighten
 
} %end tighten (references & figure captions)

\end{document}